
\tolerance=10000
\raggedbottom

\baselineskip=15pt
\parskip=1\jot

\def\sk{\vskip 3\jot}

\def\heading#1{\vskip3\jot{\noindent\bf #1}}
\def\label#1{{\noindent\it #1}}
\def\QED{\hbox{\rlap{$\sqcap$}$\sqcup$}}


\def\ref#1;#2;#3;#4;#5.{\item{[#1]} #2,#3,{\it #4},#5.}
\def\refinbook#1;#2;#3;#4;#5;#6.{\item{[#1]} #2, #3, #4, {\it #5},#6.} 
\def\refbook#1;#2;#3;#4.{\item{[#1]} #2,{\it #3},#4.}


\def\({\bigl(}
\def\){\bigr)}


\def\ga{\gamma}
\def\de{\delta}


\def\calT{{\cal T}}


\input graphicx

\def\Ex{{\rm Ex}}

\def\faslk{for all sufficiently large $k$}

{
\pageno=0
\nopagenumbers
\rightline{\tt path.search.apr.mod.tex}
\vskip1in

\centerline{\bf Local versus Global Search in Channel Graphs}
\vskip0.5in

\centerline{A. H. Hunter}
\centerline{\tt ahh@cs.washington.edu}
\centerline{Department of Computer Science \& Engineering}
\centerline{University of Washington}
\centerline{AC101 Paul G. Allen Center, Box 352350}
\centerline{185 Stevens Way}
\centerline{Seattle, WA 98195-2350}
\vskip0.5in

\centerline{Nicholas Pippenger}
\centerline{\tt njp@math.hmc.edu}
\centerline{Department of Mathematics}
\centerline{Harvey Mudd College}
\centerline{1250 Dartmouth Avenue}
\centerline{Claremont, CA 91711}
\vskip0.5in

\noindent{\bf Abstract:}
Previous studies of search in channel graphs has assumed that the search is {\it global};
that is, that the status of any link can be probed by the search algorithm at any time.
We consider for the first time {\it local\/} search, for which only links to which an idle path from the source has already been established may be probed.
We show that some well known channel graphs may require exponentially more probes, on the average, when search must be local than when it may be global.
\vfill\eject
}

\heading{1. Introduction}

A {\it channel graph\/} is an acyclic directed graph $G = (V,E)$, with {\it vertices\/} $V$ and
{\it edges\/} $E$, in which there exist a {\it source\/} vertex $s\in V$ and a {\it target\/} vertex $t\in V$
such that every vertex lies on a directed path from $s$ to $t$.
(Such a source and target, if they exist, are clearly unique.)
The vertices other than the source and target are called {\it links}.

A {\it state\/} of a channel graph is an assignment of a {\it status\/} ({\it busy\/} or {\it idle}) to each link of the graph.
We shall extend such an assignment to all vertices by agreeing that the source and target are always idle.
We shall deal in this paper with a particular probability distribution on the states of a channel graph.
We choose a real number $q$, in the range $0\le q\le 1$, which we call the {\it vacancy probability}; its complement $p = 1-q$ is called the {\it occupancy probability}.
We then define a random state of a channel graph to be one in which each link is independently idle with probability $q$ (and thus busy with probability $p$).
This probability distribution on states was introduced independently by Lee [L1] and Le Gall [L2, L3].

We shall say that a channel graph is {\it linked\/} in a given state if there exists a directed path from the source to the target consisting entirely of idle links.
We shall say that a channel graph is {\it blocked\/} in a given state if there exists a cut between the source and the target consisting entirely of busy links.
(Clearly, a channel graph in a given state is either linked or blocked, but not both.)
If a channel graph $G$ is in a random state with vacancy probability $q$,
the {\it linking probability\/} will be denoted $Q(G,q)$, and the complementary {\it blocking probability\/} will be denoted $P(G, q) = 1 - Q(G, q)$.

Consider now a search algorithm that seeks to determine whether a known channel graph,
in an unknown random state with a known vacancy probability,  is linked or blocked.
The algorithm gathers information about the state of the graph by sequentially probing the status of links until all the links of either an idle path or a busy cut have been probed.
(The algorithm may be adaptive, so that the decision as to which link to probe at a given step may depend on the outcomes of all previous probes.)

Such an algorithm may be modeled as a decision tree.
The elements of such a tree will be called {\it nodes\/} and {\it arcs\/} (to distinguish them from the vertices and edges of the channel graph).
Each node is either a {\it probe node}, in which case it is labeled with the name of a link in the channel graph and has two outgoing arcs (one labeled ``idle'' and one labeled ``busy'') leading to other nodes, or a {\it leaf}, in which case it is labeled with one of the two possible outcomes
(``linked'' or ``blocked'') and has no outgoing arcs.
There is a distinguished probe node, called the {\it root}, that has no incoming arcs; every other node has exactly one incoming arc.
Execution of the algorithm begins at the root and proceeds in an obvious way, probing links and following the appropriate arcs, until it arrives at a leaf that announces the final result.
There is an obvious notion of such an algorithm being correct:
every trajectory from the root to a leaf labeled ``linked'' probes every link on a path from the source to the target in the channel graph and departs each of these probe nodes along the ``idle'' arc,
and every trajectory from the root to a leaf labeled ``blocked'' probes every link on a cut between the source and the target in the channel graph and departs each of these probe nodes along the ``busy'' arc.
This model of a search algorithm was introduced by Lin and Pippenger [L4], and subsequently used by Pippenger [P].
We shall denote by $E(G, q)$ the minimum possible expected number of probes performed by
any search algorithm that correctly searches the channel graph $G$ with vacancy probability $q$.
This formulation of the search problem considers algorithms to be deterministic, with only the state of the network being random.
It is clear, however, that a formulation allowing randomized algorithms would yield the same function $E(G,q)$: for a randomized algorithm may be regarded as a convex combination of deterministic algorithms (that is, as a probability distribution on decision trees, obtained by making all random choices at the outset), and at least one of these algorithms must use make an expected number of probes that is at most as large as that of the convex combination.

In the definition of ``search algorithm'' given above, any node in the decision tree may probe any link in the channel graph, regardless of previous probes and their outcomes.
We shall call this situation {\it global\/} search.  
This model is appropriate for circumstances in which all parts of the network being searched
are directly accessible by the computer performing the search.
It will be unrealistic, however, if the probes themselves require communication over the network being searched.
An alternative appropriate to the latter circumstances is to allow a link to be probed only if it is  
{\it accessible}, meaning that
all the links on some path from the source to that link have previously been probed and found to be idle.
This situation will be called {\it local\/} search.

We shall denote by $E_1(G, q)$ the minimum possible expected number of probes performed by
any local search algorithm that correctly searches the channel graph $G$ with vacancy probability $q$.
We clearly have $E(G, q)\le E_1(G, q)$, and the main question explored in this paper is:
how much larger than $E(G, q)$ can $E_1(G, q)$ be?
We shall see that the answer is: for some channel graphs $G$, and some values of the vacancy 
$q$, it can be exponentially larger.
Preliminary versions of the results in this paper appeared in the first author's thesis [H2].

The graphs we shall study are called fully parallel graphs.
Let $T_k$ be a complete binary tree of depth $k$, with root $r$ and $2^k$ leaves, and with all edges directed from the root towards the leaves.
Let $T'_k$ be a similar tree of depth $k$, with root $r'$ and $2^k$ leaves, and with all edges directed from the leaves towards the root.
The {\it fully parallel\/} channel graph $F_k$ is obtained by joining each leaf of $T_k$ to the corresponding leaf of $T'_k$ by an edge directed from the former to the latter.
The source of $F_k$ is $s=r$ and the target is $t=r'$.
The vertices of $F_k$ will be partitioned into {\it ranks}: the vertices at depth $j$ ($0\le j\le k$)
of $T_k$ will constitute rank $j$; the vertices at depth $j$ ($0\le j\le k$)
of $T'_k$ will constitute rank $2k+1-j$.
Thus the source constitutes rank $0$, and the target constitutes rank
$2k+1$.  An example is shown in Figure 1.

{
\centerline{\includegraphics{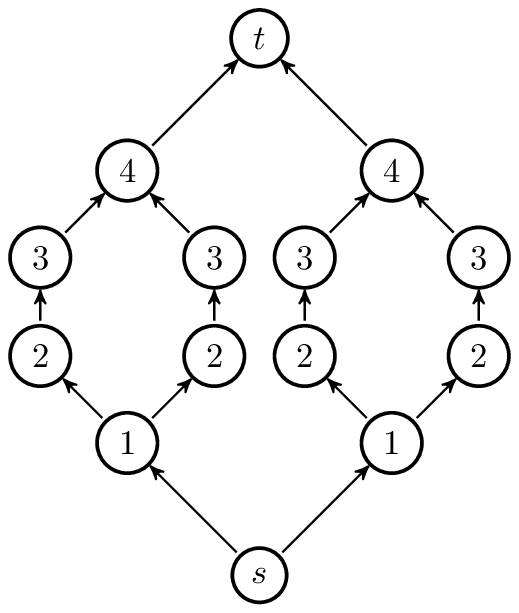}}
\smallskip
\centerline{Figure 1. The channel graph $F_2$.  Links are
annotated with their ranks.}
\smallskip
}

Our results in this paper will show that global search of $F_k$ can be performed with an expected number of probes linear in $k$ for any fixed $q$.
We shall show, however, that local search of $F_k$ requires an expected number of probes exponential in $k$ for $1/2 < q < 1$, and we shall determine the precise rate of exponential growth for $q$ in this range.

Specifically, we shall present in Section 2 an algorithm {\tt bilat-search} that performs global search in $F_k$ using an expected number of probes at most $4k$ for any fixed $k$ and \faslk.
(For $0\le q < 1/2$, the expected number of probes is in fact bounded as $k\to\infty$, but the bound depends on $q$.
Because we are primarily interested in the contrast between linear and exponential growth as a function of $k$, we shall in what follows frequently replace constants that depend on $q$ with 
factors of $k$ or $1/k$.
This replacement will weaken our results slightly, but simplify them greatly, without affecting the 
contrast between linear and exponential growth, or between different rates of exponential growth.)
We shall also present an algorithm {\tt unilat-search} that performs local search in $F_k$ using an expected number of probes at most $4k\max\{1, \min\{(2q)^k, 1/q^k\}\}$ for any fixed $q$ and \faslk.
The expected number of probes for this algorithm thus grows linearly for $0\le q\le 1/2$ and for $q=1$, but grows exponentially in $k$ for $1/2 < q < 1$, with a base that increases as $2q$ from $1$ to $\sqrt{2}$ as $q$ increases from $1/2$ to $1/\sqrt{2}$, then decreases as $1/q$ from $
\sqrt{2}$ to $1$ as $q$ increases from $1/\sqrt{2}$ to $1$.
In Section 3, we shall present lower bounds that show that this rate of exponential growth is the best possible.
Specifically, we shall show that any algorithm performing local search in $F_k$ must use an expected number of probes at least $1/kq^k$ for $1/\sqrt{2}$ and \faslk, and at least $(2q)^k/k^6$ for $1/2 < q < 1/\sqrt{2}$ and \faslk.
(The factors $1/k$ and $1/k^6$ are consequences of our desire to keep our proofs as simple as possible.
We in fact conjecture that the algorithm {\tt unilat-search} optimal not just in its rate of exponential growth, but in the stronger sense of using an expected number of probes {\it exactly\/} 
$E_1(F_k,q)$.)
\sk

\heading{2. Upper Bounds}

In this section we shall obtain upper bounds to $E(F_k, q)$ and $E_1(G, q)$ by presenting and analyzing natural path search algorithms.
The upper bound for the global case is actually a special case of a result by Lin and Pippenger [L4], but we shall give a simpler proof that can be adapted to the local case.

The results in this section and the next depend on results that are well known in the theory of branching processes (see Harris [H1]).
A {\it branching process\/} is a random process that begins with one individual in generation zero, and in which each individual independently contributes  an identically distributed random number of offspring to the next generation.
Let $Z_l$ denote the number of individuals in the $l$-th generation.
If $f(x)$ is the generating function for the number of offspring contributed by an individual,
then the generating function for $Z_l$ is the $l$-th iterate $f^{(l)}(x)$ of $f(x)$, defined by
$f^{(0)}(x) = x$ and $f^{(l+1)}(x) = f(f^{(l)}(x)) = f^{(l)}(f(x))$.

In this section we shall be concerned with the branching process for which the generating function
for the number of offspring of an individual is $f(x) = (1 - q^2 + q^2 x)^2$, describing the number of successes in two independent trials that each succeed with probability $q^2$.
This branching process governs the blocking probability $P(F_k,q)$ in the following way.
Any path from the source to the target in $F_k$ that passes through a link $v$ in $T_k$ also passes through the link $v'$ in $T'_k$, and conversely.
Consider the tree $T^*_k$ obtained from $F_k$ by identifying each vertex of $T_k$ with the corresponding vertex of $T'_k$.
The source and target of $F_k$ are identified to form the root of $T^*_k$, which we take to be idle.
Let every other vertex of $T^*_k$ be idle if and only if the corresponding links in $T_k$ and $T'_k$ are both idle (which occurs
with probability $q^2$) and busy if and only if either of the corresponding links in $T_k$ and $T'_k$ are busy (which occurs with probability $1-q^2$).
Then it is clear that $P(F_k,q)$ is equal to the probability that every path from the root to a leaf in $T^*_k$ contains at least one busy vertex.
This probability is just the probability that $Z_k = 0$, which is the constant term $f^{(k)}(0)$ in $f^{(k)}(x)$.
Thus we have the recurrence 
$$P(F_k,q) = f(P(F_{k-1},q)) \eqno(2.1)$$
for $k\ge 1$, with the initial condition $P(F_0,q) = 0$.
We have $P(F_k,q) \ge P(F_{k-1})$ for $k\ge 1$, since $Z_{k-1} = 0$ implies $Z_k = 0$.
Thus the sequence $P(F_k,q)$ is non-decreasing in $k$. 
Since it is also bounded above by $1$,
it tends to a limit, which we shall denote $P^*$, and we have $P(F_k,q)\le P^*$ for all $k\ge 0$.
Letting $k$ tend to infinity in (2.1), and noting that $f(x)$ is continuous, we see that
$P^*$ must be a fixed-point of $f(x)$, specifically the smallest fixed-point greater than or equal to $0$.
Solving the quadratic equation $f(P^*) = P^*$, we find that $P^*$ is equal to $1$ for
$0\le q\le 1/\sqrt{2}$, and equal to $(1-q^2)^2/q^4$ for $1/\sqrt{2} \le q \le 1$.
Thus we have proved the following lemma.

\label{Lemma 2.1:}
For any $0\le q\le 1$ and $k\ge 0$, we have
$$P(F_k,q) \le
\cases{
1, &if $0\le q\le {1 /  \sqrt{2}}$; \cr
& \cr
{{(1 - q^2)^2 / q^4}}, &if ${1 / \sqrt{2}}\le q\le 1$. \cr
}$$
(Note that the two cases agree for $q=1/\sqrt{2}$.)

Our upper bound for global path search in $F_k$ is based on the following recursive algorithm.
It should be clear how, for a given value of $k\ge 0$, it may be transformed into a decision tree
of the form described in the introduction.
The algorithm has been written to return {\tt true} if its argument is linked and {\tt false} if it is blocked.
It would be straightforward to add data structures that would allow it to return an idle path or busy cut as appropriate.
Since our main interest is in the cost of the search, however, we have only kept track of enough information to determine the sequence, and thus the number, of probes performed by the algorithm.
 \sk
 
{\obeylines\tt
bilat-search($F_k$ : G):
if $k=0$:
\ \ return true
let $F_{k-1}$: G', G'' be the two copies of $F_{k-1}$ in G;
if idle(source(G')) and idle(target(G')) and bilat-search(G'):
\ \ return true
if idle(source(G'')) and idle(target(G'')) and bilat-search(G''):
\ \ return true
return false
 }

\noindent
We assume short-circuiting conjunctions; that is, if {\tt f()} evaluates to
false, {\tt f() and g()} is false and  {\tt g()} is not evaluated.

From the algorithm we can easily write a recurrence for the expected number of probes it performs,
which leads to the following recurrence for $E(F_k,q)$.
We have
$$\eqalignno{
E(F_k,q)
&\le 1 + q + q^2 E(F_{k-1},q) + (p + qp + q^2P(F_{k-1},q))(1 + q + q^2 E(F_{k-1},q)) \cr
& \cr
&= (1 + p + qp + q^2P(F_{k-1},q))(1 + q + q^2 E(F_{k-1},q))  &(2.2) \cr
}$$
for $k\ge 1$, with the initial condition $E(F_0,q) = 0$.
In the right-hand side of the first line, the initial terms $1 + q +
q^2 E(F_{k-1},q)$ represents the expected cost of searching the first
copy of $F_{k-1}$, the first expression in parentheses represents the probability that no path through the first copy is found, so that the second copy must be searched, and the second expression in parentheses (which is equal to the initial terms) represents the expected cost of searching the second copy.
In the second line, we have factored out the expected cost of searching a copy, so the first expression in parentheses now represents the expected number of copies that must be searched.
Using the bound from Lemma 2.1, we obtain
$$\eqalignno{
1 + p + qp + q^2P(F_{k-1},q)
&\le 1 + p + qp + q^2P^* \cr
& \cr
& = \cases{
2, &if $0\le q\le {1 /  \sqrt{2}}$; \cr
& \cr 
1/q^2, &if ${1 / \sqrt{2}}\le q\le 1$. \cr
} &(2.3) \cr
}$$
Substituting the bounds (2.3) into the recurrence (2.2), we obtain
$$E(F_k,q) \le \cases{
2(1+q) + 2q^2 E(F_{k-1},q),  &if $0\le q\le {1 /  \sqrt{2}}$; \cr
& \cr
(1+q)/q^2  + E(F_{k-1}), &if ${1 / \sqrt{2}}\le q\le 1$ \cr
}$$
for $k\ge 1$.
Applying the bounds $2(1+q)\le 4$ and $2q^2 \le 1$ for $0\le q\le {1 /  \sqrt{2}}$ and
$(1+q)/q^2 \le 4$ for ${1 / \sqrt{2}}\le q\le 1$ yields the recurrence
$$E(F_k,q) \le 4 + E(F_{k-1},q).$$
This recurrence, together with 
the initial condition $E(F_0,q) = 0$, gives the following theorem.

\label{Theorem 2.2:}
For any $0\le q\le 1$ and $k\ge 0$, we have
$$E(F_k,q) \le 4k.$$

To obtain an upper bound for local search, we transform the above algorithm as follows.

{\obeylines\tt
unilat-search($F_k$ : G):
if $k=0$:
\ \ return true
let $F_{k-1}$: G', G'' be the two copies of $F_{k-1}$ in G;
if (idle(source(G')) and unilat-search(G')) and idle(target(G'))
\ \ return true
if (idle(source(G'')) and unilat-search(G'')) and idle(target(G''))
\ \ return true
return false
 }
 \noindent
The difference between these algorithms lies in the order of the conditions in the conjunction 
{\tt (... and ... and ...)}.  This change is necessary to keep the
 algorithm local; we may not probe {\tt target(G')} until we have
 successfully found a path through {\tt G'}, using a recursive call.
 Thus, the algorithm is much more likely to make such a recursive call
 (with probability $q$ instead of $q^2$) and its running time will
 grow much more quickly.  As before, we may read off the recurrence
$$E_1(F_k,q) \le  (1 + p + qp + q^2P(F_{k-1},q)) (1 + qE_1(F_{k-1},q) + qQ(F_{k-1},q))$$
for $k\ge 1$, with the initial condition $E_1(F_0,q) = 0$.
Using the bounds $Q(F_{k-1},q) \le 1$ and $P(F_{k-1},q) \le P^*$ together with Lemma 2.1 as before, we obtain
$$E_1(F_k,q) \le \cases{
4 + 2q E(F_{k-1},q),  &if $0\le q\le {1 /  \sqrt{2}}$; \cr
& \cr
2/q^2  + (1/q)E(F_{k-1}), &if ${1 / \sqrt{2}}\le q\le 1$ \cr
}$$
for $k\ge 1$.
Applying the bounds $2q\le 1$ for $0\le q\le 1/2$ and $2/q^2 \le 4$ for $1/\sqrt{2} \le q \le 1$ yields the recurrence
$$E_1(F_k,q) \le \cases{
4 +  E(F_{k-1},q),  &if $0\le q\le {1 / {2}}$; \cr
& \cr
4 + 2q E(F_{k-1},q),  &if $1/2 \le q\le {1 /  \sqrt{2}}$; \cr
& \cr
4  + (1/q)E(F_{k-1}), &if ${1 / \sqrt{2}}\le q\le 1$ \cr
}$$
for $k\ge 1$. 
This recurrence, together  with the initial condition $E_1(F_0,q) = 0$
yields the following theorem.

\label{Theorem 2.3:}
For any $0\le q\le 1$ and $k\ge 0$, we have
$$\eqalign{
E_1(F_k,q) &\le
\cases{
4k, &if $0\le q \le 1/2$; \cr
& \cr
4k (2q)^k, &if $1/2 \le  q \le {1 /  \sqrt{2}}$; \cr
& \cr
{4k / q^{k}}, &if ${1 / \sqrt{2}}\le q \le 1$ \cr} \cr
& \cr
&  = 4k\max\{1, \min\{(2q)^k, 1/q^k\}\}. \cr
}$$
\sk

\heading{3. Lower Bounds}

In this section, we shall prove the following theorem.

\label{Theorem 3.1:}
For $1/2 < q < 1$, we have
$$E_1(F_k,q) \ge \min\{(2q)^k / k^6, 1/kq^k\}$$
\faslk.
This result will show that the exponential rates of growth in Theorem 2.3 are the best possible.

To prove Theorem 3.1, we consider an optimal algorithm $\calT$ for local search in $F_k$ with vacancy probability $q$, and define the random variable $T$ to be the number of probes used by 
$\calT$.
Since $\calT$ is optimal, we have $\Ex[T] = E_1(F_k, q)$, so it will suffice to show that $\Ex[T]$ satisfies the lower bound of Theorem 3.1.

Our lower bounds will be based on the following principle.
If, at any point during the execution of the algorithm, we give the algorithm, at no cost,
some information that it has not asked for, that gift can only decrease the expected number of probes it needs to make to complete its task.
This principle can be formalized in terms of decision trees in the following way.
We shall transform the original decision tree $\calT$ into a modified decision tree $\calT^*$.
The tree $\calT^*$ will contain, in addition to internal nodes that make probes, internal 
{\it gift\/} nodes that branch according to the information given for free.
We shall prune the tree below each gift node, eliminating probe nodes whose outcome is determined by the gift nodes above them.
It is easy to show that this transformation preserves correctness and can only decrease the expected number of probes used:  any state of the network corresponds to a path $\pi$ from the root to a leaf $L$ in $\calT$, and a path $\pi^*$ from the root  to a leaf $L^*$ in $\calT^*$; the leaves $L$ and $L^*$ will have the same label (``linked'' or ``blocked''), and any link of the network probed by node on $\pi^*$ will also be probed by a node on $\pi$.
We shall not dwell further on this process of formalization; instead we shall proceed directly to an informal presentation of our lower bounds based on this principle.

We shall begin by telling the algorithm, before it begins its execution, the status of all accessible links in ranks $1$ through $k$, thereby informing it of which links in rank $k+1$ are accessible.
The accessible links in rank $k+1$ will be called {\it candidate\/} links.
The first probe by the algorithm will thus be to a candidate link.
Whenever the algorithm probes a candidate link, we shall tell it the status of all accessible links on the path from that candidate link to the target.
It follows that every probe by the algorithm will be to a candidate
link.  
This process is illustrated in Figure 2.
Each such probe either reveals an idle path from the probed candidate link to target (in which case the algorithm can announce ``linked''), or discovers a busy link that blocks the paths from some subset of the candidate links to the target (in which case the algorithm need not probe these candidate links,  since probing them could neither reveal an idle path to the target, nor discover a busy node that blocks any additional paths from candidate links to the target).
The algorithm will thus probe a sequence of candidate links until it either reveals an idle path to the target or discovers a set of busy links that together block the paths from all candidate links to the target.
It is clear that an optimal algorithm will never make a probe after the outcome of the search (``linked'' or ``blocked'') has been determined, nor will it probe a candidate link after a busy link has already be discovered on the path from that candidate link to the target.

{
\centerline{\includegraphics{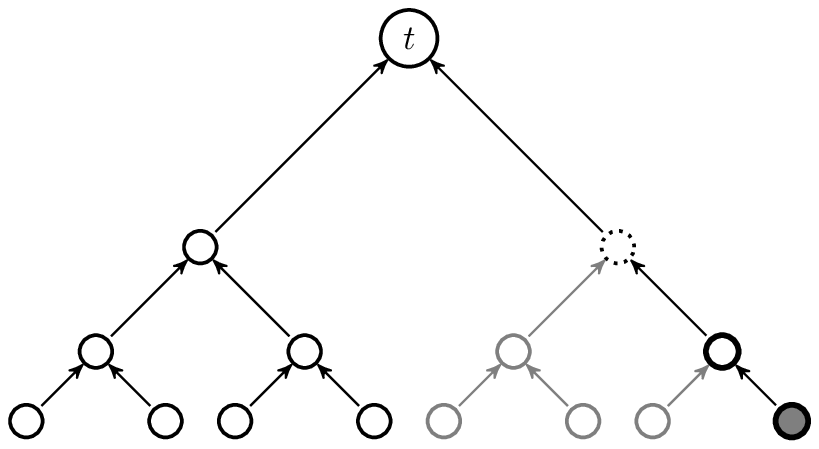}}
{\narrower\smallskip\noindent
Figure 2.  After probing the marked candidate link (gray), discovering the status of all accesible links above it (heavy for idle, dotted for busy), and pruning now-useless links (light), only candidate links are accessible.\smallskip
}
}

We can now prove the easier case, $1/\sqrt{2} < q < 1$, of Theorem 3.1.
Let $I_t$ be the event ``$T\ge t$'' (so that the algorithm performs a $t$-th probe of a candidate link), and let $J_t$ be the event ``the $t$-th probe reveals an idle path to the target'' (so that the algorithm announces ``linked'' after the $t$-th probe).
We have $\Pr[J_t \mid I_t] = q^k$ for any probe of a candidate link (since the path from any probed candidate link to the target contains $k$ links, each of which is independently idle with probability $q$).
Furthermore, by Lemma 2.1 we have $Q(F_k,q) = 1 - P(F_k,q) \ge 1 - (1-q^2)^2/q^4 \ge 1/k$ \faslk.
Thus we have
$$\eqalign{
\Ex[T]
&= \sum_{t\ge 1} \Pr[I_t] \cr
&= {1\over q^k} \sum_{t\ge 1} \Pr[J_t \mid I_t] \,\Pr[I_t] \cr
&= {1\over q^k} Q(F_k,q) \cr
&\ge {1\over kq^k} \cr
}$$
\faslk.
This proves Theorem 3.1 for the case $1/\sqrt{2} < q < 1$.

For the remaining case, $1/2 < q \le 1/\sqrt{2}$,
we shall need to work with an additional 
branching process $Y_0, Y_1, \ldots Y_l, \ldots$, for which $Y_0 = 1$ and the generating function
for the number of offspring of an individual is $g(x) = (1 - q + q x)^2$, describing the number of successes in two independent trials that each succeed with probability $q$.
This branching process governs the number of accessible links in the successive ranks of $T_k$:
the number of links in rank $1\le l\le k$ that are idle and accessible through an idle path from the source is $Y_l$.
In particular the set of candidate links has cardinality $Y_k$.
Let $A$ denote the event ``$Y_k  < (2q)^k / 2k^2$''.

\label{Lemma 3.2:}
For fixed $1/2 < q \le 1$, we have
$$\Pr[A] \le 1 - {5\over k}$$
\faslk.

\label{Proof:}
Let $M=(2q)^k/2k^2$.
Then for any $x\le 1$ we have
$$\eqalign{
\Pr[A]
&= \sum_{0\le m < M} \Pr[Y_k = m] \cr
&\le {1\over x^M}  \sum_{0\le m < M} \Pr[Y_k = m] \, x^m \cr
&\le {1\over x^M}  \sum_{0\le m \le 2^k} \Pr[Y_k = m] \, x^m \cr
&=  {1\over x^M}\, g^{(k)}(x). \cr
}$$
Let $\de = k/(2q)^k$.
Then taking $x = 1-\de$ yields
$$\Pr[A] \le {1\over (1-\de)^M}\, g{(k)}(1-\de).$$
We bound the first factor by using the inequality $(1-\de)^M \ge 1 - M\de$ (which holds for 
$0\le \de \le 1$ and $M\ge 1$), obtaining
$$\eqalign{
{1\over (1-\de)^M}
&\le {1\over 1 - (1/2k)} \cr
&= 1 + {1\over 2k-1}. \cr
}$$
It will now suffice to show that
$$g^{(k)}(1-\de) \le 1 - {6\over k}, \eqno(3.1)$$
for then we shall have
$$\eqalign{
\Pr[A] &\le \left(1 + {1\over 2k-1}\right)\left(1 - {6\over k}\right) \cr
&\le 1 - {6\over k} + {1\over 2k-1} \cr
&\le 1 - {5\over k}, \cr
}$$
completing the proof of the lemma.

The function $g(x)$ has two fixed points, at $x=1$ and at $x=(1-q)^2/q^2 < 1$.
Let $\ga = (1-q)^2/q^2$ denote this smaller fixed point.
Define $G(y) = \big(g\(\ga + (1-\ga)y\)-\ga\big)\big/\(1-\ga\) = 2(1-q)y + (2q-1)y^2$.
Since the transformation $y\mapsto \ga + (1-\ga)y$ is inverse to $x\mapsto (x-\ga)/(1-\ga)$,
iterating $G(y)$ is equivalent to iterating $g(x)$:
$g^{(k)}(x) = \ga + (1-\ga)G^{(k)}\((x-\ga)/(1-\ga)\)$.
Next define $H(y) = y/\(2q + (1 - 2q)y\)$.
Then $G(y)\le H(y)$ for all $y\ge 0$, since 
$\(2(1-q)y + (2q-1)y^2\)\(2q + (1 - 2q)y\) - y = -y(2q-1)^2 (y-1)^2 \le 0$.
Straightforward induction shows that 
$$H^{(k)}(y) = {y\over (2q)^k + (1-(2q)^k)y}.$$
Substituting $x=1-\de$ in $y=(x-\ga)/(1-\ga)$, we obtain
$y = (1-\de-\ga)/(1-\ga) < 1-\de$.
From this we obtain, since $H(y)$ and thus $H^{(k)}(y)$ are increasing functions of $y$
for $0\le y\le 1$,
$$\eqalign{
H^{(k)}(y)
&= {1-\de \over (2q)^k + (1-(2q)^k)(1-\de)} \cr
&= {1-\de \over (1-\de) + \de(2q)^k} \cr
&\le {1-\de \over (1-\de) + k} \cr
&\le {1 \over 1 + k}. \cr
}$$
Since $G^{(k)}(y) \le H^{(k)}(y)$, we obtain 
$$G^{(k)}(y) \le  {1 \over 1 + k},$$
so we have
$$\eqalign{
g^{(k)}(1-\de)
&=  \ga + (1-\ga)G^{(k)}(y) \cr
&\le \ga + {1-\ga \over 1+k}.
}$$
Since $\ga < 1$, this inequality establishes (3.1), and thus completes the proof of the lemma.
\QED

For each link $v$ in rank $k-l$ of $F_k$ for some $0\le l\le k-1$, let $X_v$
denote the number of links in rank $k$ that are accessible through idle paths from $v$.
For each link $v$ in rank $k-l$, let $B_v$ be the event $X_v > k^2 (2q)^l$'', and 
let $B$ be the event ``for some $l$ in the range $0\le l\le k-1$ and some link $v$ in rank $k-l$, 
$X_v > k^2 (2q)^l$''.

\label{Lemma 3.3:}
For fixed $1/2 < q \le 1$, we have
$$\Pr[B] \le {1\over k}$$
\faslk.

\label{Proof:}
There are at most $2\cdot 2^k$ links in $T_k$.
Thus it will suffice to prove that for any one link $v$ in rank $k-l$,
$$\Pr[B_v] \le {4\over e^k}, \eqno(3.2)$$
for then, by Boole's inequality,  we shall have
$$\eqalign{
\Pr[B]
&\le \sum_v \Pr[B_v] \cr
&\le {8\cdot 2^k \over e^k} \cr
&\le {1\over k} \cr
}$$
\faslk.

Consider a fixed link $v$ in rank $k-l$.
Let $M = k^2 (2q)^l$.
Then for any $x\ge 1$ we have
$$\eqalign{
\Pr[B_v]
&= \sum_{M<m\le 2^l} \Pr[X_v = m] \cr 
&\le {1\over x^M} \sum_{M<m\le 2^l} \Pr[X_v = m]\,x^m \cr 
&\le {1\over x^M} \sum_{0\le m\le 2^l} \Pr[X_v = m]\,x^m \cr 
&\le {1\over x^M} g^{(l)}(x). \cr
}$$ 
Let $\de = 1/k(2q)^l$.
Then taking $x = 1 + \de$ yields
$$\eqalignno{
\Pr[B_v]
&\le {1\over (1+\de)^M} \, g^{(l)}(1+\de). \cr
}$$
We bound the first factor by using the inequality $\log(1+ z) \ge z - z^2 / 2$ (which holds for all
$z\ge 0$), obtaining
$$\eqalignno{
{1\over (1+\de)^M}
&\le \exp\left(-M\log(1+\de)\right) \cr
&\le \exp\left(-M(\de + \de^2/2)\right) \cr
&= \exp\left(-k + 1/2(2q)^l\right) \cr
&\le \exp\left(-k + 1/2\right) \cr
&= {e^{1/2}\over e^k} \cr
&\le {2\over e^k}, &(3.3)\cr
}$$
\faslk.
It will now suffice to show that
$$g^{(l)}(1+\de) \le 2, \eqno(3.4)$$
\faslk,
for this combined with (3.3) will yield (3.2), completing the proof of the lemma.

Let $G(\de) = g(1+\de) - 1 = 2q\de + q^2\de^2$, and let $H(\de) = 2q\de / (1 - \de)$.
Then $G(\de) \le H(\de)$ for all $0\le \de < 1$, since
$(2q\de + q^2\de^2)(1 - \de) = 2q\de - (2 - q)q\de^2 - q^2 \de^3 \le 2q\de$.
Since $G(\de)$ and $H(\de)$ are both nondecreasing functions of $\de$ for $0\le \de < 1$,
a straightforward induction on $l\ge 0$ shows that
$$\eqalign{
G^{(l)}(\de)
&\le H^{(l)}(\de) \cr
&\le {(2q)^l \de \over 1 - l(2q)^{l-1}\de} \cr
}$$
for $0\le \de < 1/l(2q)^{l-1}$.
Taking $\de = 1/k(2q)^l$, we obtain
$$\eqalign{
g^{(l)}(1+\de)
&= 1 + G^{(l)}(\de) \cr
&\le 1 + {1/k \over 1 - 1/(2q)} \cr
&\le 2 \cr
}$$
\faslk.
This inequality establishes (3.4), and thus completes the proof of the lemma.
\QED

Let $C$ be the event ``neither $A$ nor $B$ occurs''.
From Lemmas 2.2 and 2.3 we have
$$\eqalignno{
\Pr[C]
&\ge 1 - \Pr[A] - \Pr[B] \cr
&\ge 1 - \left(1 - {5\over k}\right) - {1\over k} \cr
&= {4\over k}. &(3.5) \cr
}$$

As before, let $I_t$ denote the event ``$T\ge t$''.
If the $t$-th probe of a candidate link discovers $l$ accessible idle links on the path from the candidate link to the target (before reaching a busy link or the target), we shall associate with 
with that probe a ``payoff'' $K_t = k^2 \, (2q)^l$.
The number $L_t$ of accessible idle links discovered on the path from the probed candidate link to the target has the distribution $\Pr[L_t = l\mid I_t] = (1-q)q^l$ for $0\le l\le k-1$ and
$\Pr[L_t=k\mid I_t] = q^k$, and thus the payoff $K_t$ of the $t$-th probe has the distribution
$\Pr[K_t=k^2\,(2q)^l\mid I_t] = (1-q)q^l$ for $0\le l\le k-1$ and $\Pr[K_t=k^2\,(2q)^k\mid I_t] = q^k$.
We thus have
$$\eqalign{
\Ex[K_t\mid I_t]
&\le k^2 \, \sum_{0\le l\le k} (2q)^l \, q^l \cr
&\le 2k^3, \cr
}$$
since $2q^2 \le 1$.
We note that these distributions and expectations are independent of $A$, $B$ and $C$.

Let $K = \sum_{y\ge 1} K_t$ denote the total payoff from all probes.
If $C$ occurs, and after $T$ probes of candidate links the algorithm announces ``linked'' or ``blocked'', the total payoff from all probes of candidate links must satisfy
$K \ge (2q)^k/2k^2$.
For if the algorithm announces ``blocked'', then (because $A$ did not occur) there must have been at least $(2q)^k/2k^2$ candidate links, and (because $B$ did not occur)
the number of candidate links whose paths to the target were found to be blocked by the $t$-th probe was at most the payoff $K_t$ (since the candidate links whose paths to the target blocked by a busy link in rank $k+1+l$ are accessible through idle paths from a link in rank $k-l$).
And if the algorithm announced ``linked'', the $T$-th probe must have revealed an idle path from a candidate link to the target, and this probe alone had payoff $K_T = k^2 \, (2q)^k \ge (2q)^k / 2k^2$.

Thus we have
$$\eqalign{
\Ex[T]
&\ge \Ex[T\mid C] \, \Pr[C] \cr
&\ge {1\over k} \Ex[T\mid C] \cr
&\ge {1\over k} \sum_{t\ge 1} \Pr[I_t\mid C] \cr
&\ge {1\over k}\,{1\over 2k^3} \sum_{t\ge 1} \Ex[K_t\mid I_t]\,\Pr[I_t\mid C] \cr
&\ge {1\over k}\,{1\over 2k^3} \sum_{t\ge 1} \Ex[K\mid C] \cr
&\ge  {1\over k}\,{1\over 2k^3}\,{(2q)^k \over 2k^2} \cr
&= {(2q)^k\over 4k^6}. \cr
}$$ 
This completes the proof of Theorem 3.1 in the case $1/2 < q \le 1/ \sqrt{2}$.
\sk

\heading{4. Conclusion}

We have presented a sequence $F_k$ of channel graphs for which global path search is easy,
in the sense that its cost is $O(k)$ for any vacancy probability $q$, as is shown by the natural bilateral depth-first search algorithm presented in Section 2.
We have shown in Section 3 that the cost of local path search in $F_k$ is exponential in $k$ for all $1/2 < q < 1$.
We have also presented in Section 2 a natural algorithm, unilateral depth-first search, that shows that the exponential rate of growth in shown in Section 3 is the best possible, in the sense that it matches the exponential rate of growth of the cost of unilateral depth-first search.
We conjecture that unilateral depth-first search is in fact optimal in the stronger sense of making the smallest possible expected number of probes for every $k\ge 0$ and $0\le q\le 1$.

The foregoing results may be summarized by saying that global search in $F_k$ is cheap, but local search is expensive.
If, however, we consider ``bilateral local search'', wherein a link may be probed if either there is an idle path from the source to it, or an idle path from it to the target, then search in $F_k$ is again cheap,
because the global search algorithm {\tt bilat-search} probes only vertices meeting one of these conditions.

Are there channel graphs in which global search is cheap, but even bilateral local search is expensive?
The answer is ``yes'', as can be seen by considering the graph $F_k \circ F_k$ obtained by connecting two copies of $F_k$ in series, with the target of the first copy identified with the source of the second copy to form a link $u$.
Global search of $F_k \circ F_k$ is cheap:
we first probe $u$;  if it is idle, we then search the first copy of $F_k$; and if that copy is linked, we search the second copy.
Thus
$$E(F_k \circ F_k, q) \le 1 + q(1 + Q(F_k, q))E(F_k, q).$$
If, however, we condition on the link $u$ being idle, and give this information for free to the search algorithm,
then a path in $F_k \circ F_k$ is the concatenation of two paths, one in each copy of $F_k$, while a cut in $F_k \circ F_k$ must include a cut in at least one copy of $F_k$.
Since $F_k$ is symmetric under reversal, every link probed by a bilateral local search in $F_k \circ F_k$
can be probed in a ``unilateral'' local search of the first copy of $F_k$, or in a ``reverse unilateral''
local search of the second copy.
Thus the lower bound of Section 3 for unilateral search in $F_k$, when multiplied by the 
probability $q$ that $u$ is idle, becomes a lower bound for bilateral search in $F_k \circ F_k$.
Thus even bilateral local search of $F_k \circ F_k$ is expensive.
\sk

\heading{5. Acknowledgment}

The research reported in this paper was supported in part by Grant CCF 0646682 from the National Science Foundation.
\sk

\heading{6. References}


\refbook H1; T. E. Harris;
The Theory of Branching Processes;
Springer-Verlag, 1963.

\refbook H2; A. H. Hunter;
Locality \& Complexity in Path Search;
B.~S. Thesis, Department of Mathematics, 
Harvey Mudd College, May 2009.


\ref L1; C. Y. Lee;
``Analysis of Switching Networks'';
Bell System Technical Journal; 34 (1955) 1287--1315.

\ref L2; P. Le Gall;
``\'{E}tude du blocage dans les syst\`{e}mes de commutation t\'{e}l\'{e}phonique automatique
utilisant des commutateurs \'{e}lectroniques du type crossbar'';
Ann.\  des T\'{e}l\'{e}comm.; 11 (1956) 159--171; 180--194; 197.

\ref L3; P. Le Gall;
````M\'{e}thode de calcul de l'encombrement dans les syst\`{e}mes t\'{e}l\'{e}phonique automatique
\`{a} marquage'';
Ann.\  des T\'{e}l\'{e}comm.; 12 (1957) 374--386.

\ref L4; G. Lin and N. Pippenger;
``Routing Algorithms for Switching Networks with Probabilistic Traffic'';
Networks; 28:1 (1996) 21--29.

\ref P;  N. Pippenger;
``Upper and Lower Bounds for the Average-Case Complexity of Path Search'';
Networks; 33:4 (1999) 249--259.

\bye